\documentstyle[aps,multicol,epsf]{revtex}

\begin{document}
\title{\large\bf
The First Search for Gravitational Waves from 
Inspiraling Compact Binaries using TAMA300 data 
}
\author{
Hideyuki  Tagoshi$^{  1}$,
Nobuyuki  Kanda$^{  2}$,
Takahiro  Tanaka$^{  3}$,
Daisuke   Tatsumi$^{  4}$,
Souichi   Telada$^{  5}$,
Masaki    Ando$^{  6}$,
Koji      Arai$^{  5}$,
Akito     Araya$^{   7}$,
Hideki    Asada$^{  8}$,
Mark A.   Barton$^{  4}$,
Masa-Katsu Fujimoto$^{  5}$,
Mitsuhiro Fukushima$^{  7}$,
Toshifumi Futamase$^{   9}$,
Gerhard Heinzel$^{  5}$,
Gen'ichi  Horikoshi$^{  10}$\footnote{deceased},
Hideki    Ishizuka$^{  4}$,
Norihiko  Kamikubota$^{   10}$,
Keita     Kawabe$^{  6}$,
Seiji     Kawamura$^{  5}$,
Nobuki    Kawashima$^{   11}$,
Yasufumi  Kojima$^{  12}$,
Yoshihide Kozai$^{  5}$,
Kazuaki   Kuroda$^{  4}$,
Namio     Matsuda$^{  13}$,
Sumihiro  Matsumura$^{  4}$,
Satoshi   Miki$^{  14}$,
Norikatsu Mio$^{  14}$,
Osamu     Miyakawa$^{  4}$,
Shoken    Miyama$^{  5}$,
Shinji    Miyoki$^{  4}$,
Eiichi    Mizuno$^{  15}$,
Shigenori Moriwaki$^{  14}$,
Mitsuru   Musha$^{  16}$,
Shigeo    Nagano$^{  4}$,
Ken'ichi  Nakagawa$^{  16}$,
Takashi   Nakamura$^{  3}$,
Ken-ichi  Nakao$^{  17}$,
Kenji     Numata$^{  6}$,
Yujiro    Ogawa$^{   10}$,
Masatake  Ohashi$^{  4}$,
Naoko     Ohishi$^{  6}$,
Akira     Okutomi$^{ 4}$,
Ken-ichi  Oohara$^{  18}$,
Shigemi   Otsuka$^{  6}$,
Yoshio    Saito$^{   10}$,
Misao     Sasaki$^{  1}$,
Shuichi   Sato$^{  4}$,
Atsushi   Sekiya$^{  6}$,
Masaru    Shibata$^{  1}$,
Kazumichi Shirakata$^{  14}$,
Kentaro   Somiya$^{  14}$,
Toshikazu Suzuki$^{   10}$,
Ryutaro   Takahashi$^{  5}$,
Akiteru   Takamori$^{  6}$,
Shinsuke  Taniguchi$^{  6}$,
Kuniharu  Tochikubo$^{  6}$,
Takayuki  Tomaru$^{  4}$,
Kimio     Tsubono$^{  6}$,
Nobuhiro  Tsuda$^{  19}$,
Takashi   Uchiyama$^{  4}$,
Akitoshi  Ueda$^{  5}$,
Ken-ichi  Ueda$^{  16}$,
Kozo      Ueda$^{  6}$,
Koichi    Waseda$^{  5}$,
Yuko      Watanabe$^{  2}$,
Hiromi    Yakura$^{  2}$,
Kazuhiro  Yamamoto$^{  6}$, and 
Toshitaka Yamazaki$^{  5}$\\
(The TAMA Collaboration)
}
\address{\it
{${}^{1}$Department of Earth and Space Science, Osaka University,
Toyonaka, Osaka 560-0043, Japan
}\\
{${}^{2}$Department of Physics, Miyagi University of Education,
Aoba Aramaki, Sendai 980-0845, Japan
}\\
{${}^{3}$Yukawa Institute for Theoretical Physics, Kyoto University
}\\
{${}^{4}$Institute for Cosmic Ray Research, The University of Tokyo
}\\
{${}^{5}$National Astronomical Observatory of Japan
}\\
{${}^{6}$Department of Physics, The University of Tokyo
}\\
{${}^{7}$Earthquake Research Institute, The University of Tokyo 
}\\
{${}^{8}$Faculty of Science and Technology, Hirosaki University
}\\
{${}^{9}$Graduate School of Science, Tohoku University
}\\
{${}^{10}$High Energy Accelerator Research Orginization
}\\
{${}^{11}$Kinki University
}\\
{${}^{12}$Department of Physics, Hiroshima University
}\\
{${}^{13}$Tokyo Denki University
}\\
{${}^{14}$Department of Advanced Materials Science, The University of Tokyo
}\\
{${}^{15}$The Institute of Space and Astronautical Science
}\\
{${}^{16}$Institute for Laser Science, University of Electro-Communications
}\\
{${}^{17}$Department of Physics, Osaka City University
}\\
{${}^{18}$Faculty of Science, Niigata University
}\\
{${}^{19}$Precision Engineering Division, Faculty of Engineering, Tokai University
}
}
\address{\vspace*{5mm}
\centerline{\rm 
\begin{minipage}{16cm}\hspace*{5mm}
We analyzed 6 hours of data from the TAMA300 detector
by matched filtering, searching for gravitational waves from
inspiraling compact binaries.
We incorporated a two-step hierarchical search strategy in matched filtering. 
We obtained an upper limit of 0.59/hour (C.L.=90$\%$) 
on the event rate of inspirals of compact 
binaries with mass between 0.3$M_\odot$ and 10$M_\odot$ and 
with signal-to-noise ratio greater than 7.2. 
The distance of 1.4$M_\odot$ (0.5$M_\odot$) binaries 
which produce the signal-to-noise ratio 7.2 was
estimated to be 6.2kpc (2.9kpc) when the position of 
the source on the sky and the inclination angle 
of the binaries were optimal.
\end{minipage}}
}
\maketitle

\begin{multicols}{2}
{\it Introduction}: 
Several laser interferometric gravitational wave detectors 
are now under construction. 
These include LIGO\cite{ref:LIGO}, VIRGO\cite{ref:VIRGO}, 
GEO600\cite{ref:GEO}, and TAMA300\cite{ref:TAMA}. 
The TAMA300 detector has been developed 
over the past five years. 
It is a power-recycled, 
Fabry-Perot-Michelson interferometer, 
which consists of mirrors that are suspended
by vibration isolation systems.
The differential armlength changes in the 300m Fabry-Perot arm cavities
caused by propagation of gravitational waves are 
monitored by the Michelson interferometer.

The TAMA300 detector became ready to operate in the summer of 1999.
Most of the designed systems  (except
power recycling) were installed by that time.
The first test operation was done on August 6th 1999. 
The first long-term data were taken for three nights
between September 17th and 20th 1999
\cite{ref:TAMA}\cite{ref:ando}. 
The total data length amounted to about 30 hours, with
the longest continuous lock time of the interferometer lasting nearly 8 hours.
The strain equivalent noise spectrum is given in Fig. 1. 
The best sensitivity was about $3\times 10^{-20}/\sqrt{{\rm Hz}}$ 
around 900Hz. 
The data of the first two days contained 
many burst noises due to the instability of the interferometer system 
and it was very difficult to use those data for the gravitatioal wave
event search. However, 
the quality of the data of the final day was good enough to perform 
the search for gravitational waves. 

Currently, TAMA300 is the world's largest detector 
in operation. 
This is the first time that 
long continuous data from one of the 
large-scale interferometer of the new generation quoted above were taken. 
By analyzing the data, 
we expect to obtain useful knowledge of the property of 
long and continuous data. 
It is also important to develop tools for analyzing such data 
as a first step to treat the much larger amount of data 
which will be obtained in the near future. 
This is the first paper on data analysis of TAMA300, 
in which we report the first 
result of our search of gravitational waves from 
inspiraling compact binaries. 

{\it Inspiraling binaries}: 
Gravitational waves from inspiraling compact binaries,
consisting of neutron stars with mass $\sim 1.4M_\odot$
or black holes, 
have been considered to be the most promising target for 
laser interferometers.
These compact binaries can be produced as a consequence of 
normal stellar evolution of binaries. 
It has been also suggested that 
MACHOs\cite{ref:macho} in our Galactic halo may be primordial black holes
with mass $\sim 0.5M_\odot$. 
If so, it is reasonable to 
expect that some of them are in binaries which 
coalesce due to the gravitational radiation reaction\cite{ref:bhmacho}. 

Inspiraling compact binaries are nearly ideal sources 
for data analysis, since 
their waveforms can be theoretically calculated to a very high
accuracy. However, in this paper, we did not take account of
the effect of spin angular momentum. Thus, we cannot deny the 
possibility that we lose signals from binaries with spin.
\cite{ref:apostolatos}.

{\it Data Acquisition and Calibration}: 
The main signals from the detector were derived from the feedback 
signal that keeps the interferometer in resonance, which
was digitized by an 
analog-to-digital converter in 16 bit depth with 20kHz sampling rate.
The noise amplitude due to quantization by the ADC was estimated
to be $10^{-2} - 10^{-3}$ 
times smaller than the instrumental (electronic) noise.
A precise sampling clock was generated using a GPS-derived 10MHz
reference signal.
All of the data were recorded in the ``Common Data Frame
Format\cite{ref:frame}''
on tape archives. 

The strain of a gravitational wave
$h(t)$ and the voltage signal $V(t)$ are related in the frequency domain by
$\tilde{h}(f)=\tilde{F}(f) \tilde{V}(f)$, 
where $\tilde{F}(f)$ is a response function. 
We measured the full open-loop response function in
the observed frequency range before and after each
continuous operation. During the observation,  we
continuously monitored drifts of the response
function by adding a single frequency sinusoidal
signal into the feedback system and observing its
amplitude and phase at different points of the loop\cite{ref:telada}.
This method has an advantage that the observation is not interrupted
by the calibration signal. 
The accuracy of strain measurement was evaluated to 
$\Delta h/h \simeq 1\%$ within the observed frequency band. 
Using Monte-Carlo simulations, we also confirmed that
systematic errors in the outputs of matched filter 
caused by calibration errors were negligible 
compared with noise fluctuations.
The details of this method will be presented in
a separate paper\cite{ref:telada2}. 

{\it Matched filtering}:
We denote the strain equivalent 
one-sided power spectrum density of the noise by $S_n(f)$.
In order to calculate the expected wave forms, 
which are called {\it templates},  
we used restricted post-Newtonian
wave forms of order 2.5, in which the phase evolution was correctly taken
into account up to the 2.5 post-Newtonian order, but 
the amplitude was evaluated by using the quadrupole formula. 
As for the 2.5 post-Newtonian phase evolution, 
we used formulas derived by Blanchet et al.\cite{ref:pntemplate}. 

When the gravitational wave passes through the interferometer, 
it produces a relative difference $\Delta L$ between the  
two armlengths $L$. The gravitational wave strain amplitude 
is defined by $\Delta L=L h(t)$. 
The wave form $h(t)$ is calculated by combining two independent modes 
of the gravitational wave and the antenna pattern of the interferometer as 
\begin{equation}
h(t)={\cal A}[h_c(t-t_c)~\cos\alpha +h_s(t-t_c)~\sin\alpha ],
\end{equation}
where $t_c$ is the coalescence time, and 
$h_c(t)$ and $h_s(t)$ are the two independent templates with 
the phase difference $\pi/2$. 
To construct filters, we need 
the Fourier transforms of $h_{c}(t)$ and $h_{s}(t)$. 
They were computed directly by using the stationary phase 
approximation, the validity of which 
was established by Droz et al. \cite{ref:DKPO}. 
The parameters to distinguish the wave forms are 
the amplitude ${\cal A}$, the two masses $m_1, m_2$, 
the coalescence time $t_c$ and the phase $\alpha$. 
We did not include spins of the stars 
in the parameters. 

We denote the data from the detector as $s(t)$. 
We define a filtered unnormalized signal-to-noise ratio $\rho$ after 
the maximization over $\alpha$ as 
\begin{eqnarray}
\rho&=&\sqrt{(s,h_c)^2+(s,h_s)^2}, \label{eq:defrho} \\
(a,b)&\equiv&2\int^{\infty}_{-\infty}df
{\tilde{a}(f)\tilde{b}^*(f)\over S_n(|f|)},
\label{eq:innerproduct}
\end{eqnarray}
where $\tilde{a}(f)$ denotes the Fourier transform of $a(t)$
and the asterisk denotes the complex conjugation. 
This $\rho$ has an expectation value $\sqrt{2}$
in the presence of only Gaussian noise. 
Thus, the signal-to-noise ratio, SNR, 
is given by SNR$=\rho/\sqrt{2}$. 

Analyzing the real data from TAMA300, we found 
that the noise contained a large amount of non-stationary 
and non-Gaussian noise whose statistical properties have not 
been understood well yet.
In order to remove the influence of such noise, 
we introduced a $\chi^2$ test of 
the time-frequency behavior of the signal \cite{ref:40m}. 
We divide each template into $n$ mutually independent pieces
in the frequency domain,
chosen so that the expected contribution to $\rho$ from 
each frequency band is approximately equal. 
For two template polarizations $h_{c}(t)$ and $h_{s}(t)$, 
we calculate $\chi^2$ by summing the square 
of the deviation of each value of $\rho$ from the expected 
value\cite{ref:grasp}. 
This quantity must satisfy the $\chi^2$-statistics with
$2n-2$ degrees of freedom, 
as long as the data consists of Gaussian noise plus 
chirp signals. 
However, there was a 
strong tendency that an event with large $\chi^2$ has 
a large value of $\rho$. 
Thus, by applying a threshold to the $\chi^2$ value, 
we can reduce the number of fake events without 
significantly losing the detectability of real events. 
For convenience, we renormalized $\chi^2$ as 
$\chi^2/(2n-2)$. 
In the current analysis, we chose $n=16$. 
This number was determined mainly by the limited memory
of our computer. 

We searched the parameter space of $0.3M_\odot$ 
$\leq$ $m_1, m_2$ $\leq$ $10M_\odot$ with total mass less than
$10M_\odot$. 
The low mass limit was chosen so that it covers the
estimated mass of MACHOs as much as possible. 
In this parameter space, we prepared a mesh. 
The mesh points define the templates used for search. 
The spacing of the mesh points was determined 
so as not to lose more than 2 $\%$ of signal-to-noise 
due to the mismatch between actual mass parameters 
and those at mesh points. 
Using geometrical arguments, we introduced a new 
parameterization of masses that simplifies the algorithm 
to determine the mesh points\cite{ref:TT}. 

The parameter space defined by using our new mass parameters
turned out to contain 2057 templates in the present analysis. 
Although this number 
is not too large to perform matched filtering with a simple algorithm,
when the noise power spectrum of data is improved, 
the necessary number of templates will increase. 
Thus, we introduced a two-step hierarchical search algorithm 
and developed tools of matched filtering which can be used to analyze 
future much larger data streams.

{\it two-step hierarchical search}:
The basic idea of the two-step search is simple. 
The templates for the first step search are chosen  
at the mesh points with coarse spacing. 
We perform the first step search 
with an appropriate threshold which must be chosen to be 
sufficiently low not to lose real events. 
When there are events that
exceed the threshold, we search the points 
around them in detail. 

Before starting the two-step search, 
we evaluated the values of $\rho$ and $\chi^2$ of all the data
with a few selected templates 
to obtain an estimate of the background 
distribution of $\rho$ and $\chi^2$, 
assuming all of these events are due to noise.
Based on this distribution, we looked for several combinations of 
$(\rho^{*},\chi^{*2})$  
so that there were only a few fake events which satisfy  
$\rho>\rho^*$ and $\chi^2<\chi^{*2}$ simultaneously. 
These thresholds became our preliminary second step thresholds.
The first step threshold and the first step length of the mesh separation 
were determined to satisfy the following condition (C1):
{\it When $\rho$ and $\chi^2$ of an event
at the second step mesh point satisfies a second step threshold, 
such an event must also be detected at, at least, one of the 
first step mesh points around the second step point 
with more than 98 $\%$ probability}. 
By imposing this condition, it is guaranteed that 
the false dismissal rate of real detectable events 
due to the presence of the first step search
is kept to be small\cite{ref:mohanty}. 
Note that the preliminary second step thresholds, 
$(\rho^{*},\chi^{*2})$, which were used to decide 
the first step thresholds, determined the detection probability of 
events with given amplitude. 
These preliminary second step threshold are given in Table I(a). 

We performed simulations by adding signals to real data, 
and looked for combinations of the first step thresholds and 
the mesh separation which satisfy the criterion (C1). 
We chose one of them which minimizes the 
computation time in total. 
We introduced, at the first step, the $\chi^2$ threshold as well as 
the $\rho$ threshold to select the candidates for the second step, 
which reduced the computation time for the second step. 

In order to reduce the computation time, we further introduced 
various techniques, some of which are explained in \cite{ref:TT}. 

Here, we mention the cut off frequency $f_c$ of templates. 
The post-Newtonian wave forms must be terminated at some frequency
roughly corresponding to
the inner most stable circular orbit, 
at which the binaries are expected 
to transit to plunge orbit to coalesce.
However, it is difficult to determine, theoretically, 
the optimal value of $f_c$. 
At the first step search, the choice of $f_c$ is important 
because the mesh separation is large. 
If the signal was located 
at a point far from the first step mesh points, 
the value of $\chi^2$ could become much larger than unity 
due to the difference of $f_c$ between the first step template and 
the signal. 
Thus, at the first step search, 
we searched for the value of $f_c$ which produced 
the largest value of $\rho$. 
On the other hand, at the second step search, 
we simply adopted $f_c=c^3/(7^{3/2}\pi GM)$
where $M$ is the total mass, 
$G$ is the Newton's constant, and $c$ is the speed of light. 
This corresponds to the orbital radius of $r\sim 7GM/c^2$.

{\it Results}:
Our analysis was done with 8 Compaq Alpha machines.
Each of them can perform the FFT 
of data consisting of $\sim 10^{6}$ 
single precision real numbers with 140MFlops. 

{}From the 30 hours of total data, 
we selected the data with better quality taken between
14:42 and 21:04 19th September 1999 (UTC time).
The data consist of two separate stretches when
the interferometer was continuously locked.
An interval of 12.5 minutes around an unlocked part 
was not used for matched filtering. 
The total length of data used for matched filtering
was $6.1663$ hours. 

In Fig.\ref{fig:D_M_SNR}, 
we show a contour plot of the signal-to-noise ratio 
as a function of the distance to the source and 
the total mass for equal mass binaries. 
We found that the signal-to-noise ratio becomes maximum around 
1.6$M_\odot$ binaries and decreases above 
this mass. This is because the noise level of the detector increases 
rapidly below several hundreds Hz. 

In performing the two-step search,
the data were divided into data segments 
of 265.42 seconds with an overlap of
29.9 seconds, which was longer than the longest template.
The strain power spectrum density $S_n(f)$ was estimated 
using 530.84  seconds of data near the segment which was analyzed. 

As a result of the second step search, we obtain $\rho$ and $\chi^2$ 
as functions of masses and the coalescing time $t_c$. 
In each small interval of coalescing time $\Delta t_c$, 
we looked for an event which had the maximum $\rho$ 
and which satisfies certain $\chi^2$ threshold. 
In Fig.\ref{fig:fitting}, we show the numbers of these events 
which survived after the maximization
as a function of SNR for various $\chi^2$ thresholds. 
The time interval $\Delta t_c$ was chosen so that
the numbers of correlated events were small enough, but at the same time 
there was a sufficiently large number of events to 
investigate the properties of its distribution. 
In Fig.\ref{fig:fitting}, we chose $\Delta t_c=2$ seconds. 

The computation time needed to perform 
the two-step search depends strongly on the first step threshold. 
If we had optimized the first step threshold by applying 
the above condition (C1), we would have obtained the threshold given in
Table I(b) in the case when the unit area of the parameter space 
of the first step search
is 72 times larger than that of the second step. 
In that case, the computation time would have been
about 1.5 hours. 
However, we used a lower first step threshold as given in Table I(c)
to obtain the result of Fig.\ref{fig:fitting}, 
in order to investigate the background 
distribution which was used for statistical analysis.
The computation time thus needed to obtain the results
in Fig.\ref{fig:fitting} was about 5.5 hours. 
In this analysis, we used a fixed set of the first step thresholds and spacing 
for all of the data. Thus, the number of events which satisfy
the first step threshold, and therefore total computation time, 
differs very much among the portion of data. 

Using this result, 
we estimate an upper limit on the event rate. 
In Fig.\ref{fig:fitting}, 
we show a fitting to the number of events as a function 
of SNR$^{2}$. 
The value of the $\chi^2$ threshold, $1.5$, was chosen so that 
there was a $3.8\%$ chance of rejecting real events in Gaussian
noise. 
We chose the SNR threshold as SNR$=7.2$.
Thus, the observed number of events which exceed this threshold is 2. 
Using this analytic fitting, we estimate the expected number 
of background events $N_{BG}$ which is larger than SNR$=7.2$
as $N_{BG}=2.5$. 
Thus, using Bayesian statistics, and assuming 
uniform prior probability for the real event rate and 
the Poisson distribution of 
real and background events, 
we estimate 
with $90\%$ confidence that the expected number of real events
which exceed SNR$=7.2$ 
is smaller than 3.68 in this data\cite{ref:basian}.
Thus, we obtained the upper limit of the event rate 
as 0.59/hour (C.L.$=90\%$). 
Note that we did not assume that the
observed 2 events were real or not. 
(The largest SNR event for $\chi^2<1.5$ 
has SNR=9.0 and $\chi^2=1.35$. Although this event seems to show 
a small excess from the background distribution for $\chi^2<1.5$,
it also seems to be absorbed in the background for $\chi^2<2.5$. )

{}From Fig.2, we find the distance of a source that would produce
${\rm SNR}=7.2$ as $2.9$kpc for $m_1=m_2=0.5M_\odot$, and
$6.2$kpc for $m_1=m_2=1.4M_\odot$ respectively.
In this paper, we estimate the distance 
in the case when the position of 
the source on the sky and the inclination angle 
of the binaries were optimal. 
In order to evaluate the efficiency, 
we performed simulations. 
We added artificial signals to the data and performed the same
two-step search which derived Fig.\ref{fig:fitting}.
We then evaluated the detection probability of artificial signals
with the final thresholds determined in Fig.\ref{fig:fitting}. 
The result is given in Fig.\ref{fig:efficiency}. 
We found that TAMA300 could observe 1.4$M_\odot$ events in several kpc.
In other words, TAMA300 with the present sensitivity 
can give a direct observational limit of the coalescing rate 
within several kpc.
Here, it is important to note again that 
a new detector, TAMA300, produces data which are good enough 
to perform data analysis and to give upper limit of the event rate, 
even if it is sensitive only to events within several kpc. 

The sensitivity of TAMA300 detector is now being improved rapidly. 
With its goal sensitivity, 
$1.4M_\odot$ ($0.5M_\odot$)  binaries
at distance $D=820$kpc ($D=340$kpc)
will produce a signal-to-noise ratio SNR=10.
It is planned that much longer data with improved 
sensitivity and stability will be taken during the year 2000. 
By analyzing such data, we will be able to obtain a much more 
stringent upper limit to the event rate. 
And if we are very lucky, we may find a plausible candidate for a 
real gravitational wave signal. 

This work was supported 
by Monbusho Grant-in-Aid for Creative Basic Research 09NP0801. 
This work was also supported in part by 
Monbusho Grant-in-Aid 11740150, 12640269, 
and by Research Fellowships of the Japan Society 
for the Promotion of Science for Young Scientists.

\begin{figure}[h]
\epsfysize= 4.5cm 
\centerline{\epsfbox{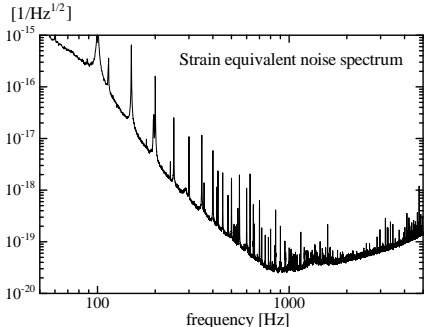}} 
\caption{The strain equivalent 
noise spectrum of TAMA300 on September 19th.
The best sensitivity is 
about $3\times 10^{-20}/\sqrt{{\rm Hz}}$ 
around 900Hz.}
\label{fig:noise}
\end{figure}

\begin{figure}[h]
\epsfysize= 4.5cm 
\centerline{\epsfbox{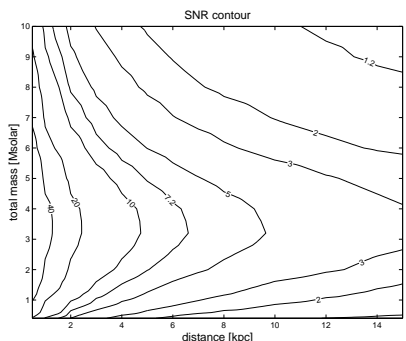}} 
\caption{A contour plot of SNR as a function of 
distance and total mass of equal mass binaries 
in the case when the position of 
the source on the sky and the inclination angle 
of the binaries are optimal.}
\label{fig:D_M_SNR}
\end{figure}

\begin{figure}[h]
\epsfysize= 4.5cm 
\centerline{\epsfbox{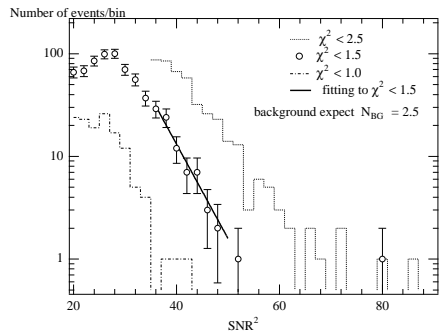}} 
\caption{The number of events for $\chi^2<2.5$, 
$\chi^2<1.5$ and $\chi^2<1.0$ as a function of SNR$^2$, 
and an analytic fitting to $\chi^2<1.5$. 
The fitting was determined between SNR$^2$=35 and 50. 
This fitting gives the number of events larger than 
SNR=7.2 and $\chi^2<1.5$ as 2.5. 
The $\chi^2<2.5$, $\chi^2<1.5$, and 
$\chi^2<1.0$ correspond to $10^{-3}\%$, $3.8\%$, 
and 46$\%$ false dismissal rate in Gaussian noise, respectively.}
\label{fig:fitting}
\end{figure}

\begin{figure}[h]
\epsfysize= 4.5cm 
\centerline{\epsfbox{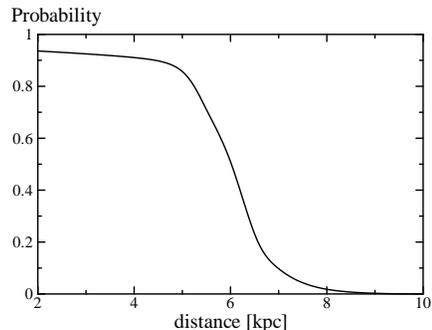}} 
\caption{The detection probability of artificial signals of 
1.4$M_\odot$ binaries as a function of distance in the case 
when the final thresholds were (SNR$^*$,$\chi^{*2}$)= (7.2,1.5). }
\label{fig:efficiency}
\end{figure}

\begin{table}
\caption{
(a) The preliminary second step thresholds used to 
determine the first step thresholds and the mesh spacing.
(b) Optimized first step thresholds. (c) The first step thresholds 
used to derive the result of Fig.3 and 4.
In both cases, the first step mesh area is 72 times larger than 
that of the second step. }
\begin{center}
\begin{minipage}{5cm}
\begin{tabular}{cc|c||c|c|c|c}
&\multicolumn{2}{c||}{(a)} &
\multicolumn{2}{c|}{(b)} & \multicolumn{2}{c}{(c)} \\ \tableline
&$\rho^*$ & $\chi^{*2}$ &
$\rho^*_{\rm 1st}$ & $\chi^{*2}_{\rm 1st}$ &
$\rho^*_{\rm 1st}$ & $\chi^{*2}_{\rm 1st}$ \\ \tableline
&9.3  & 1.0 &8.0  & 1.3 & 7.0 & 1.3 \\ \tableline
&10.1 & 1.3 &8.2  & 1.5 & 7.2 & 1.5 \\ \tableline
&10.6 & 1.5 &8.5  & 2.0 & 7.5 & 2.0 \\ \tableline
&11.4 & 1.7 &9.0  & 2.5 & 8.0 & 2.5 \\ \tableline
&12.1 & 2.0 &10.2  & 3.0 & 9.2 & 3.0 \\ \tableline
&13.7 & 2.5 &&&&
\end{tabular}
\end{minipage}
\end{center}
\end{table}

\end{multicols}

\end{document}